\documentclass[runningheads]{llncs}
\usepackage{graphicx}
\usepackage[dvipsnames]{xcolor}

\usepackage[utf8]{inputenc}
\usepackage{amsmath}

\definecolor{Su}{RGB}{0, 77, 154}
\definecolor{In}{RGB}{154, 0, 77}

\definecolor{Re}{RGB}{85,168,104}

\definecolor{DarkRed}{RGB}{132,0,0}

\usepackage{amsfonts}
\usepackage{graphicx}
\usepackage{color}
\usepackage{subfig}
\usepackage[labelfont=bf]{caption}

\usepackage{ wasysym }
\usepackage{tikz}
\usetikzlibrary{positioning,calc}
\usepackage{color}

\usepackage[misc]{ifsym} 

\usepackage{mathtools}
\usepackage{dirtytalk}
\usepackage[frak=mma]{mathalfa}

\usepackage{bbold}
\usepackage{hyperref}

\usepackage{amssymb}

\usepackage{comment}
\usepackage{scalerel}

\newcommand{\posInt}{\mathbb{Z}_{\geq 0}}
\newcommand{\vecm}{\mathbf{m}}
\newcommand{\Vecm}{\mathcal{M}}

\newcommand{\vecv}{\mathbf{v}}
\newcommand{\Vecv}{\mathcal{V}}
\newcommand{\kmax}{k_{\text{max}}}
\newcommand{\suml}[1]{\sum\limits_{#1}}

\newenvironment{thisnote}{\par\color{black}}{\par}

\newcommand{\gerrit}[1]{\textcolor{red}{GG:#1}}
\newcommand{\revision}[1]{\textcolor{black}{#1}}

\DeclarePairedDelimiterX\Set[1]{\{}{\}}{
  
  #1
}

\newcommand{\oStates}{\mathcal{S}}

\begin{document}
\title{Reducing Spreading Processes on Networks to Markov Population Models}
%
%
\author{Gerrit Großmann\textsuperscript{(\Letter)}\inst{1}\orcidID{0000-0002-4933-447X} \and
Luca Bortolussi\inst{1,2}\orcidID{0000-0001-8874-4001} 
}
\authorrunning{G. Großmann and L. Bortolussi}
%
\institute{Saarland University, 66123 Saarbrücken, Germany
\email{gerrit.grossmann@uni-saarland.de}
\and
University of Trieste, Trieste, Italy \\
\email{lbortolussi@units.it}
}
\maketitle              
\begin{abstract}
Stochastic processes on complex networks, where each node is in one of several compartments, and neighboring nodes interact with each other, can be used to describe a variety of real-world spreading phenomena. 
However, computational analysis of such processes is hindered by the enormous size of their underlying state space. 

In this work, we demonstrate that lumping can be used to reduce any epidemic model to a Markov Population Model (MPM). Therefore, we propose a novel lumping scheme based on a partitioning of the nodes. By imposing different types of counting abstractions, we obtain coarse-grained Markov models with a natural MPM representation that approximate the original systems.
This makes it possible to transfer the rich pool of approximation techniques developed for MPMs to the computational analysis of complex networks' dynamics. 



We present numerical examples to investigate the relationship between the accuracy of the MPMs, the size of the lumped state space, and the type of counting abstraction.

\keywords{Epidemic Modeling \and Markov Population Model \and Lumping \and Model Reduction \and Spreading Process \and SIS Model \and Complex Networks}
\end{abstract}
\section{Introduction} 
\raggedbottom
Computational modeling and analysis of dynamic processes on networked systems is a wide-spread and thriving research area. In particular, much effort has been put into the study of spreading phenomena \cite{barabasi2016network,porter2016dynamical,goutsias2013markovian,kiss2016mathematics}. 
Arguably, the most common formalism for spreading processes is the so-called \texttt{S}usceptible-\texttt{I}nfected-\texttt{S}usceptible ($\texttt{SIS}$) model with its variations \cite{kiss2016mathematics,porter2016dynamical,rodrigues2016application}.

In the $\texttt{SIS}$ model, each node is either \emph{infected} ($\texttt{I}$) or \emph{susceptible} ($\texttt{S}$). Infected nodes propagate their infection to neighboring susceptible nodes and become susceptible again after a random waiting time. 
Naturally, one can extend the number of possible node states (or compartments) of a node. For instance, the $\texttt{SIR}$ model introduces an additional \emph{recovered} state in which nodes are immune to the infection.

$\texttt{SIS}$-type models are remarkable because---despite their simplicity---they allow the emergence of complex macroscopic phenomena guided by the topological properties of the network.
There exists a wide variety of scenarios which can be described using the $\texttt{SIS}$-type formalism.
For instance, the $\texttt{SIS}$ model has been successfully used to study the spread of many different pathogens like influenza \cite{keeling2011modeling}, dengue fever \cite{rodrigues2010dynamics}, and SARS \cite{ng2003double}. 
Likewise, $\texttt{SIS}$-type models have shown to be extremely useful for analyzing and predicting the spread of opinions \cite{watts2007influentials,kitsak2010identification},
rumors \cite{zhao2012sihr,zhao2013sir}, and memes \cite{wei2013competing} in online social networks. Other areas of applications include the modeling of neural activity \cite{goltsev2010stochastic}, the spread of computer viruses \cite{gan2012propagation} 
as well as blackouts in financial institutions \cite{may2009systemic}.

The semantics of $\texttt{SIS}$-type processes can be described using a continuous-time Markov chain (CTMC) \cite{kiss2016mathematics,van2009virus} (cf. Chapter \ref{Formal} for details). Each possible assignment of nodes to the two node states $\texttt{S}$ and $\texttt{I}$ constitutes an individual state in the CTMC (here referred to as \emph{network state} to avoid confusion\footnote{In the following, we will use the term CTMC state and network state interchangeably.}). Hence, the CTMC state space grows exponentially with the number of nodes, which renders the numeral solution of the CTMC infeasible for most realistic contact networks. 

This work investigates an aggregation scheme that \emph{lumps} similar network states together and thereby reduces the size of the state space. 
More precisely, we first partition the nodes of the contact network. 
After which, we impose a counting abstraction on each partition. 
We only lump two networks states together when their corresponding counting abstractions coincide on each partition. 

As we will see, the counting abstraction induces a natural representation of the lumped CTMC as a Markov Population Model (MPM). In an MPM, the CTMC states are vectors which, for different types of species, count the number of entities of each species. The dynamics can elegantly be represented as species interactions. More importantly, a very rich pool of approximation techniques has been developed on the basis of MPMs, which can now be applied to the lumped model. These include efficient simulation techniques \cite{cao2006efficient,allen2009efficient},
dynamic state space truncation \cite{henzinger2009sliding,mateescu2010fast}, moment-closure approximations \cite{soltani2015conditional,grima2012study}, 
linear noise approximation \cite{van1992stochastic,grima2010}, and hybrid approaches \cite{bortolussi2016,singh2010stochastic}.


The remainder of this work is organized as follows: Section \ref{related} shortly revises related work, Section \ref{Formal} formalized $\texttt{SIS}$-type models and their CTMC semantics. Our lumping scheme is developed in Section \ref{ApproximateLumping}. In Section \ref{MPM}, we show that the lumped CTMCs have a natural MPM representation.  
Numerical results are demonstrated in in Section \ref{casestudies} and some conclusions in Section \ref{conclusion} complete the paper and identify open research problems.

\section{Related Work\label{related}}
The general idea behind \emph{lumping} is to reduce the complexity of a system by aggregating (i.e., lumping) individual components of the system together.
Lumping is a popular model reduction technique which has been used to reduce the number of equations in a system of ODEs and the number of states in a Markov chain, in particular in the context of biochemical reaction networks \cite{li1990general,buchholz1994exact,wei1969lumping,cardelli2017erode}. 
Generally speaking, one can distinguish between \emph{exact} and \emph{approximate} lumping \cite{li1990general,buchholz1994exact}.

Most work on the lumpability of epidemic models has been done in the context of exact lumping \cite{kiss2016mathematics,simon2011exact,ward2018general}. The general idea is typically to reduce the state space by identifying symmetries in the CTMC which themselves can be found using symmetries (i.e., automorphisms) in the contact network.
Those methods, however, are limited in scope because these symmetries are infeasible to find in real-world networks and the state space reduction is not sufficient to make realistic models small enough to be solvable. 

This work proposes an approximate lumping scheme.
Approximate lumping has been shown to be useful when applied to mean-field approximation approaches of epidemic models like the degree-based mean-field and pair approximation equations \cite{kyriakopoulos2018lumping}, as well as the approximate master equation \cite{grossmann2018lumping,gleeson2013binary}. 
However, mean-field equations are essentially inflexible as they do not take topological properties into account or make unrealistic independence assumptions between neighboring nodes. 

Moreover, \cite{khudabukhsh2018approximate} proposed using local symmetries in the contact network instead of automorphisms to construct a lumped Markov chain.
This scheme seems promising, in particular on larger graphs where automorphisms often do not even exist, however, the limitations for real-world networks due to a limited amount of state space reduction and high computational costs seem to persist.

Conceptually similar to this work is also the \emph{unified mean-field framework} (UMFF) proposed by Devriendt et al.\;in \cite{devriendt2017unified}. Devriendt et al.\;also partition the nodes of the contact network but directly derive a mean-field equation from it. In contrast, this work focuses on the analysis of the lumped CTMC and its relation to MPMs. Moreover, we investigate different types of counting abstractions, not only node based ones. 

\section{Spreading Processes\label{Formal}}
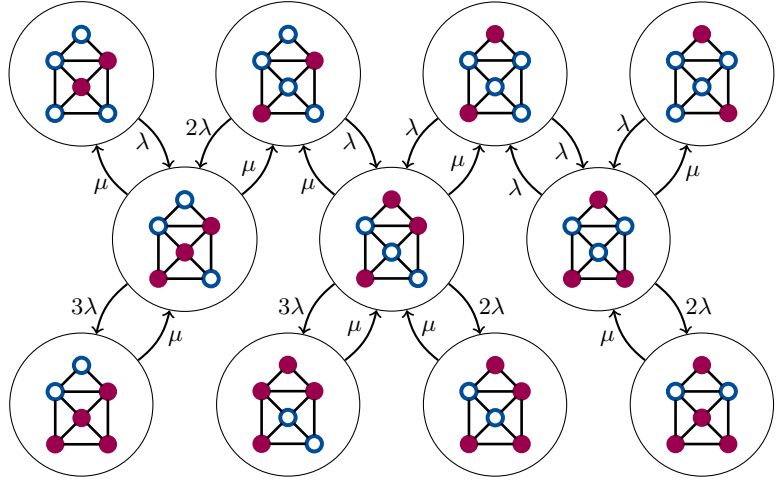
\begin{figure}[t!]
    \centering
    \begin{tikzpicture}[scale=1.1,remember picture,
    inner/.style={circle,draw=blue!50,fill=blue!20,thick,inner sep=3pt},
    outer/.style={circle}
    ]
    \node[circle,draw] (A) at (0,0) {
        \begin{tikzpicture}[scale=0.35, transform shape]
        \node (a) at (0,0) [circle ,draw , In, line width=1.5pt,fill=In] {I};
        \node (b) at (0,2) [circle ,draw , Su, line width=1.5pt] {\phantom{S}};
        \node (c) at (2,2) [circle ,draw , In, line width=1.5pt,fill=In] {I};
        \node (d) at (2,0) [circle ,draw , Su, line width=1.5pt] {\phantom{S}};
        \node (e) at (1,1) [circle ,draw , In, line width=1.5pt,fill=In] {I};
        \node (f) at (1,3) [circle ,draw , Su, line width=1.5pt] {\phantom{S}};
        \draw[line width=1.0pt] (a) edge (b) (b) edge (c) (b) edge (e) (e) edge (a) (e) edge (c) (a) edge (d)  (d) edge (c)  (c) edge (f) (b) edge (f) (d) edge (e);
        \end{tikzpicture}
    };
    \node[circle,draw] (B) at (2.5,0) {
        \begin{tikzpicture}[scale=0.35, transform shape]
        \node (a) at (0,0) [circle ,draw , In, line width=1.5pt,fill=In] {I};
        \node (b) at (0,2) [circle ,draw , Su, line width=1.5pt] {\phantom{S}};
        \node (c) at (2,2) [circle ,draw , In, line width=1.5pt,fill=In] {I};
        \node (d) at (2,0) [circle ,draw , Su, line width=1.5pt] {\phantom{S}};
        \node (e) at (1,1) [circle ,draw , Su, line width=1.5pt] {\phantom{S}};
        \node (f) at (1,3) [circle ,draw , In, line width=1.5pt,fill=In] {I};
        \draw[line width=1.0pt] (a) edge (b) (b) edge (c) (b) edge (e) (e) edge (a) (e) edge (c) (a) edge (d)  (d) edge (c)  (c) edge (f) (b) edge (f) (d) edge (e);
        \end{tikzpicture}
    };
    \node[circle,draw] (C) at (5,0) {
        \begin{tikzpicture}[scale=0.35, transform shape]
        \node (a) at (0,0) [circle ,draw , In, line width=1.5pt,fill=In] {I};
        \node (b) at (0,2) [circle ,draw , Su, line width=1.5pt] {\phantom{S}};
        \node (c) at (2,2) [circle ,draw , Su, line width=1.5pt] {\phantom{S}};
        \node (d) at (2,0) [circle ,draw , In, line width=1.5pt,fill=In] {I};
        \node (e) at (1,1) [circle ,draw , Su, line width=1.5pt] {\phantom{S}};
        \node (f) at (1,3) [circle ,draw , In, line width=1.5pt,fill=In] {I};
        \draw[line width=1.0pt] (a) edge (b) (b) edge (c) (b) edge (e) (e) edge (a) (e) edge (c) (a) edge (d)  (d) edge (c)  (c) edge (f) (b) edge (f) (d) edge (e);
        \end{tikzpicture}
    };
    \node[circle,draw] (D) at (-1.25, -2.0) {
        \begin{tikzpicture}[scale=0.35, transform shape]
        \node (a) at (0,0) [circle ,draw , In, line width=1.5pt,fill=In] {I};
        \node (b) at (0,2) [circle ,draw , Su, line width=1.5pt] {\phantom{S}};
        \node (c) at (2,2) [circle ,draw , In, line width=1.5pt,fill=In] {I};
        \node (d) at (2,0) [circle ,draw , In, line width=1.5pt,fill=In] {I};
        \node (e) at (1,1) [circle ,draw , In, line width=1.5pt,fill=In] {I};
        \node (f) at (1,3) [circle ,draw , Su, line width=1.5pt] {\phantom{S}};
        \draw[line width=1.0pt] (a) edge (b) (b) edge (c) (b) edge (e) (e) edge (a) (e) edge (c) (a) edge (d)  (d) edge (c)  (c) edge (f) (b) edge (f) (d) edge (e);
        \end{tikzpicture}
    };
    \node[circle,draw] at (1.25, -2.0) (E) {
        \begin{tikzpicture}[scale=0.35, transform shape]
        \node (a) at (0,0) [circle ,draw , In, line width=1.5pt,fill=In] {I};
        \node (b) at (0,2) [circle ,draw , In, line width=1.5pt,fill=In] {I};
        \node (c) at (2,2) [circle ,draw , In, line width=1.5pt,fill=In] {I};
        \node (d) at (2,0) [circle ,draw , Su, line width=1.5pt] {\phantom{S}};
        \node (e) at (1,1) [circle ,draw , Su, line width=1.5pt] {\phantom{S}};
        \node (f) at (1,3) [circle ,draw , In, line width=1.5pt,fill=In] {I};
        \draw[line width=1.0pt] (a) edge (b) (b) edge (c) (b) edge (e) (e) edge (a) (e) edge (c) (a) edge (d)  (d) edge (c)  (c) edge (f) (b) edge (f) (d) edge (e);
        \end{tikzpicture}
    };
    \node[circle,draw] at (3.75, -2.0) (F) {
        \begin{tikzpicture}[scale=0.35, transform shape]
        \node (a) at (0,0) [circle ,draw , In, line width=1.5pt,fill=In] {I};
        \node (b) at (0,2) [circle ,draw , Su, line width=1.5pt] {\phantom{S}};
        \node (c) at (2,2) [circle ,draw , In, line width=1.5pt,fill=In] {I};
        \node (d) at (2,0) [circle ,draw , In, line width=1.5pt,fill=In] {I};
        \node (e) at (1,1) [circle ,draw , Su, line width=1.5pt] {\phantom{S}};
        \node (f) at (1,3) [circle ,draw , In, line width=1.5pt,fill=In] {I};
        \draw[line width=1.0pt] (a) edge (b) (b) edge (c) (b) edge (e) (e) edge (a) (e) edge (c) (a) edge (d)  (d) edge (c)  (c) edge (f) (b) edge (f) (d) edge (e);
        \end{tikzpicture}
    };
    \node[circle,draw] at (6.25, -2.0) (G) {
        \begin{tikzpicture}[scale=0.35, transform shape]
        \node (a) at (0,0) [circle ,draw , In, line width=1.5pt,fill=In] {I};
        \node (b) at (0,2) [circle ,draw , Su, line width=1.5pt] {\phantom{S}};
        \node (c) at (2,2) [circle ,draw , Su, line width=1.5pt] {\phantom{S}};
        \node (d) at (2,0) [circle ,draw , In, line width=1.5pt,fill=In] {I};
        \node (e) at (1,1) [circle ,draw , In, line width=1.5pt,fill=In] {I};
        \node (f) at (1,3) [circle ,draw , In, line width=1.5pt,fill=In] {I};
        \draw[line width=1.0pt] (a) edge (b) (b) edge (c) (b) edge (e) (e) edge (a) (e) edge (c) (a) edge (d)  (d) edge (c)  (c) edge (f) (b) edge (f) (d) edge (e);
        \end{tikzpicture}
    };
    \node[circle,draw] at (-1.25, 2.0) (H) {
        \begin{tikzpicture}[scale=0.35, transform shape]
        \node (a) at (0,0) [circle ,draw , Su, line width=1.5pt] {\phantom{S}};
        \node (b) at (0,2) [circle ,draw , Su, line width=1.5pt] {\phantom{S}};
        \node (c) at (2,2) [circle ,draw , In, line width=1.5pt,fill=In] {I};
        \node (d) at (2,0) [circle ,draw , Su, line width=1.5pt] {\phantom{S}};
        \node (e) at (1,1) [circle ,draw , In, line width=1.5pt,fill=In] {I};
        \node (f) at (1,3) [circle ,draw , Su, line width=1.5pt] {\phantom{S}};
        \draw[line width=1.0pt] (a) edge (b) (b) edge (c) (b) edge (e) (e) edge (a) (e) edge (c) (a) edge (d)  (d) edge (c)  (c) edge (f) (b) edge (f) (d) edge (e);
        \end{tikzpicture}
    };
    \node[circle,draw] at (1.25, 2.0) (I) {
        \begin{tikzpicture}[scale=0.35, transform shape]
        \node (a) at (0,0) [circle ,draw , In, line width=1.5pt,fill=In] {I};
        \node (b) at (0,2) [circle ,draw , Su, line width=1.5pt] {\phantom{S}};
        \node (c) at (2,2) [circle ,draw , In, line width=1.5pt,fill=In] {I};
        \node (d) at (2,0) [circle ,draw , Su, line width=1.5pt] {\phantom{S}};
        \node (e) at (1,1) [circle ,draw , Su, line width=1.5pt] {\phantom{S}};
        \node (f) at (1,3) [circle ,draw , Su, line width=1.5pt] {\phantom{S}};
        \draw[line width=1.0pt] (a) edge (b) (b) edge (c) (b) edge (e) (e) edge (a) (e) edge (c) (a) edge (d)  (d) edge (c)  (c) edge (f) (b) edge (f) (d) edge (e);
        \end{tikzpicture}
    };
    \node[circle,draw] at (3.75, 2.0) (J) {
        \begin{tikzpicture}[scale=0.35, transform shape]
        \node (a) at (0,0) [circle ,draw , In, line width=1.5pt,fill=In] {I};
        \node (b) at (0,2) [circle ,draw , Su, line width=1.5pt] {\phantom{S}};
        \node (c) at (2,2) [circle ,draw , Su, line width=1.5pt] {\phantom{S}};
        \node (d) at (2,0) [circle ,draw , Su, line width=1.5pt] {\phantom{S}};
        \node (e) at (1,1) [circle ,draw , Su, line width=1.5pt] {\phantom{S}};
        \node (f) at (1,3) [circle ,draw , In, line width=1.5pt,fill=In] {I};
        \draw[line width=1.0pt] (a) edge (b) (b) edge (c) (b) edge (e) (e) edge (a) (e) edge (c) (a) edge (d)  (d) edge (c)  (c) edge (f) (b) edge (f) (d) edge (e);
        \end{tikzpicture}
    };
    \node[circle,draw] at (6.25, 2.0) (K) {
        \begin{tikzpicture}[scale=0.35, transform shape]
        \node (a) at (0,0) [circle ,draw , Su, line width=1.5pt] {\phantom{S}};
        \node (b) at (0,2) [circle ,draw , Su, line width=1.5pt] {\phantom{S}};
        \node (c) at (2,2) [circle ,draw , Su, line width=1.5pt] {\phantom{S}};
        \node (d) at (2,0) [circle ,draw , In, line width=1.5pt,fill=In] {I};
        \node (e) at (1,1) [circle ,draw , Su, line width=1.5pt] {\phantom{S}};
        \node (f) at (1,3) [circle ,draw , In, line width=1.5pt,fill=In] {I};
        \draw[line width=1.0pt] (a) edge (b) (b) edge (c) (b) edge (e) (e) edge (a) (e) edge (c) (a) edge (d)  (d) edge (c)  (c) edge (f) (b) edge (f) (d) edge (e);
        \end{tikzpicture}
    };
    \draw[->,line width=0.8pt]  (A)  to [bend right=20] node[left, pos=0.5]{$3 \lambda$} (D);
    \draw[->,line width=0.8pt]  (D)  to [bend right=20] node[right, pos=0.5]{$\mu$} (A);
    \draw[->,line width=0.8pt]  (B)  to [bend right=20] node[left, pos=0.5]{$3 \lambda$} (E);
    \draw[->,line width=0.8pt]  (E)  to [bend right=20] node[left, pos=0.7]{$\mu$} (B);
    \draw[->,line width=0.8pt]  (B)  to [bend left=20] node[right, pos=0.5]{$2 \lambda$} (F);
    \draw[->,line width=0.8pt]  (F)  to [bend left=20] node[right, pos=0.7]{$\mu$} (B);
    \draw[->,line width=0.8pt]  (C)  to [bend left=20] node[right, pos=0.5]{$2 \lambda$} (G);
    \draw[->,line width=0.8pt]  (G)  to [bend left=20] node[left, pos=0.5]{$\mu$} (C);
    \draw[->,line width=0.8pt]  (H)  to [bend left=20] node[left, pos=0.5]{$\lambda$} (A);
    \draw[->,line width=0.8pt]  (A)  to [bend left=20] node[left, pos=0.2]{$\mu$} (H);
    \draw[->,line width=0.8pt]  (I)  to [bend right=20] node[left, pos=0.2]{$2 \lambda$} (A);
    \draw[->,line width=0.8pt]  (A)  to [bend right=20] node[left, pos=0.6]{$\mu$} (I);
    \draw[->,line width=0.8pt]  (I)  to [bend left=20] node[left, pos=0.5]{$\lambda$} (B);
    \draw[->,line width=0.8pt]  (B)  to [bend left=20] node[left, pos=0.2]{$\mu$} (I);
    \draw[->,line width=0.8pt]  (J)  to [bend right=20] node[left, pos=0.2]{$\lambda$} (B);
    \draw[->,line width=0.8pt]  (B)  to [bend right=20] node[left, pos=0.7]{$\mu$} (J);
    \draw[->,line width=0.8pt]  (J)  to [bend left=20] node[left, pos=0.7]{$\lambda$} (C);
    \draw[->,line width=0.8pt]  (C)  to [bend left=20] node[left, pos=0.2]{$\lambda$} (J);
    \draw[->,line width=0.8pt]  (K)  to [bend right=20] node[left, pos=0.1]{$\lambda$} (C);
    \draw[->,line width=0.8pt]  (C)  to [bend right=20] node[right, pos=0.5]{$\mu$} (K);
\end{tikzpicture}
\caption[CTMC State Space of SIS Model]{The CTMC induced by the $\texttt{SIS}$ model ($\texttt{S}$: \emph{blue}, $\texttt{I}$: \emph{magenta, filled}) on a toy graph. Only a subset of the CTMC spate space ($11$ out of $2^6=64$ network states) is shown.}
\label{SISCTMC}
\end{figure}

Let $\mathcal{G}=(\mathcal{N}, \mathcal{E})$ be a an undirected graph without self-loops.
At each time point $t \in \mathbb{R}_{\geq 0}$ each node occupies one of $m$ different node states, denoted by $\mathcal{S} = \{s_1, s_2, \dots, s_m\}$ (typically, $\mathcal{S}=\{\mathtt{S},\mathtt{I}\} )$.
Consequently, the network state is given by a labeling $x: \mathcal{N} \rightarrow \mathcal{S}$.
We use 
\begin{equation*}
    \mathcal{X} = \{x \mid x: \mathcal{N} \rightarrow \mathcal{S} \}
\end{equation*}
to denote all possible labelings. $\mathcal{X}$ is also the state space of the underlying CTMC. As each of the $|\mathcal{N}|$ nodes occupies one of $m$ states, we find that $|\mathcal{X}| = |\mathcal{S}|^{|\mathcal{N}|}$.

A set of stochastic rules determines the particular way in which nodes change their corresponding node states. Whether a rule can be applied to a node depends on the state of the node and of its immediate neighborhood.

The neighborhood of a node is modeled as a vector $\mathbf{m} \in \mathbb{Z}_{\geq0}^{|\mathcal{S}|}$ where $\mathbf{m}[s]$ denotes the number of neighbors in state $s \in \mathcal{S}$ (we assume an implicit enumeration of states).
Thus, the degree (number of neighbors, denoted by $k$) of a node is equal to the sum over its associated neighborhood vector, that is, $k=\sum_{s \in \mathcal{S}} \mathbf{m}[s]$.
The set of possible neighborhood vectors is denoted as 
\begin{equation*}
\Vecm=\Big\{ \vecm \in \posInt^{|\mathcal{S}|} \mathrel{\bigg|} \sum_{s\in \mathcal{S}} \vecm[\text{s}] \leq \kmax \Big\} \;,
\end{equation*}
where $\kmax$ denotes the maximal degree in a given network. 

Each rule is a triplet $s_1 \xrightarrow{f} s_2 $ ($s_1,s_2 \in \mathcal{S}, s_1 \neq s_2$), which can be applied to each node in state $s_1$.
When the rule \say{fires} it transforms the node from $s_1$ into $s_2$. The rate at which a rule \say{fires} is specified by the rate function $f: \mathcal{M} \rightarrow \mathbb{R}_{\geq 0}$ and depends on the node's neighborhood vector. 
The time delay until the rule is applied to the network state is drawn from an exponential distribution with rate $f(\vecm)$. Hence, higher rates correspond to shorter waiting times.
For the sake of simplicity and without loss of generality, we assume that for each pair of states $s_1$, $s_2$ there exists at most one rule that transforms $s_1$ to $s_2$.

In  the  well-known $\texttt{SIS}$ model,  infected  nodes propagate their infection to susceptible neighbors. Thus, the rate at which a susceptible node becomes infected is proportional to its number of infected neighbors:
\begin{equation*}
 \phantom{XXXX} 
\mathtt{S} \xrightarrow{f} \mathtt{I} \phantom{XX}  \text{with} \phantom{X} f(\vecm) = \lambda \cdot \vecm[\mathtt{I}] \;,
\end{equation*}
where $\lambda \in \mathbb{R}_{\geq 0}$ is a rule-specific rate constant (called \emph{infection rate}) and $\vecm[\mathtt{I}]$ denotes the number of infected neighbors. 
Furthermore, a recovery rule transforms infected nodes back to being susceptible:
\begin{equation*}
 \phantom{XXXX} 
\mathtt{I} \xrightarrow{f} \mathtt{S} \phantom{XX}  \text{with} \phantom{X} f(\vecm) = \mu\;, \phantom{\cdot \vecm[\mathtt{I}]} 
\end{equation*}
where $\mu \in \mathbb{R}_{\geq 0}$ is a rule-specific rate constant called \emph{recovery rate}.

A variation of the $\texttt{SIS}$ model is the $\texttt{SI}$ model where no curing rule exists and all nodes (that are reachable from an infected node) will eventually end up being infected.  
Intuitively, each rule tries to \say{fire} at each position $n\in \mathcal{N}$ where it can be applied. The rule and node that have the shortest waiting time \say{win} and the rule is applied there. This process is repeated until some stopping criterion is fulfilled.

\subsection{CTMC Semantics\label{CTMCsemantics}}

Formally, the semantics of the $\texttt{SIS}$-type processes can be given in terms of continuous-time Markov Chains (CTMCs). The state space is the set of possible network states $\mathcal{X}$. The CTMC has a transition from state $x$ to $x'$ ($x,x' \in \mathcal{X}$, $x\neq x'$) if there exists a node $n \in \mathcal{N}$ and a rule $s_1 \xrightarrow{f} s_2 $ such that the application of the rule to $n$ transforms the network state from $x$ to $x'$. The rate of the transition is exactly the rate $f(\vecm)$ of the rule when applied to $n$. 
We use $q(x,x') \in \mathbb{R}_{\geq 0}$ to denote the transition rate between two network states. Fig.~\ref{SISCTMC} illustrates the CTMC corresponding to an $\texttt{SIS}$ process on a small toy network.

Explicitly computing the evolution of the probability of $x \in \mathcal{X}$ over time with an ODE solver, using numerical integration, is only possible for very small contact networks, since the state space grows exponentially with the number of nodes.
Alternative approaches include sampling the CTMC, which can be done reasonably efficiently even for comparably large networks \cite{grossmann2018rejection,cota2017optimized,st2019efficient} but is subject to statistical inaccuracies and is mostly used to estimate global properties.

\section{Approximate Lumping\label{ApproximateLumping}}
Our lumping scheme is composed of three basic ingredients:
\smallskip 

\noindent\textbf{Node Partitioning}: 
The partitioning over the nodes $\mathcal{N}$ that is explicitly provided. \\
\noindent\textbf{Counting Pattern}: The type of features we are counting, i.e., nodes or edges. \\
\noindent\textbf{Implicit State Space Partitioning}:  The CTMC state space is implicitly partitioned by counting the nodes or edges on each node partition.

\smallskip

We will start our presentation discussing the partitioning of the state space, then showing how to obtain it from a given node partitioning and counting pattern. To this end, we use $\mathcal{Y}$ to denote the new \emph{lumped} state space and assume that there is a surjective\footnote{If $\mathcal{L}$ is not surjective, we consider only the image of $\mathcal{L}$ to be the lumped state space. } lumping function 
\begin{equation*}
    \mathcal{L}: \mathcal{X} \rightarrow \mathcal{Y}
\end{equation*}
that defines which network states will be lumped together. Note that the lumped state space is the image of the lumping function and that all network states $x \in \mathcal{X}$ which are mapped to the same $y \in \mathcal{Y}$ will be aggregated.

Later in this section, we will discuss concrete realizations of $\mathcal{L}$. 
In particular, we will construct $\mathcal{L}$ based on a node partitioning and a counting abstraction of our choice.
Next, we define the the transition rates $q(y,y')$ (where $y,y' \in \mathcal{Y}$, $y \neq y'$) between the states of the lumped Markov chain:
\begin{equation}
\label{transition}
    q(y,y') = \frac{1}{|\mathcal{L}^{-1}(y)|}  \phantom{.}
    \suml{x \in \mathcal{L}^{-1}(y)} \phantom{.} \suml{x' \in \mathcal{L}^{-1}(y')} q(x,x')\;.
\end{equation}
This is simply the mean transition rate at which an original state from $x$ goes to some $x'\in \mathcal{L}^{-1}(y')$.
Technically, Eq.~\eqref{transition} corresponds to the following \emph{lumping assumption}: we assume that at each point in time all network states belonging to a lumped state $y$ are equally likely. 

\subsection{Partition-Based Lumping\label{lumping}}

\begin{figure}%
    \centering
    \subfloat[]{{\includegraphics[width=6cm]{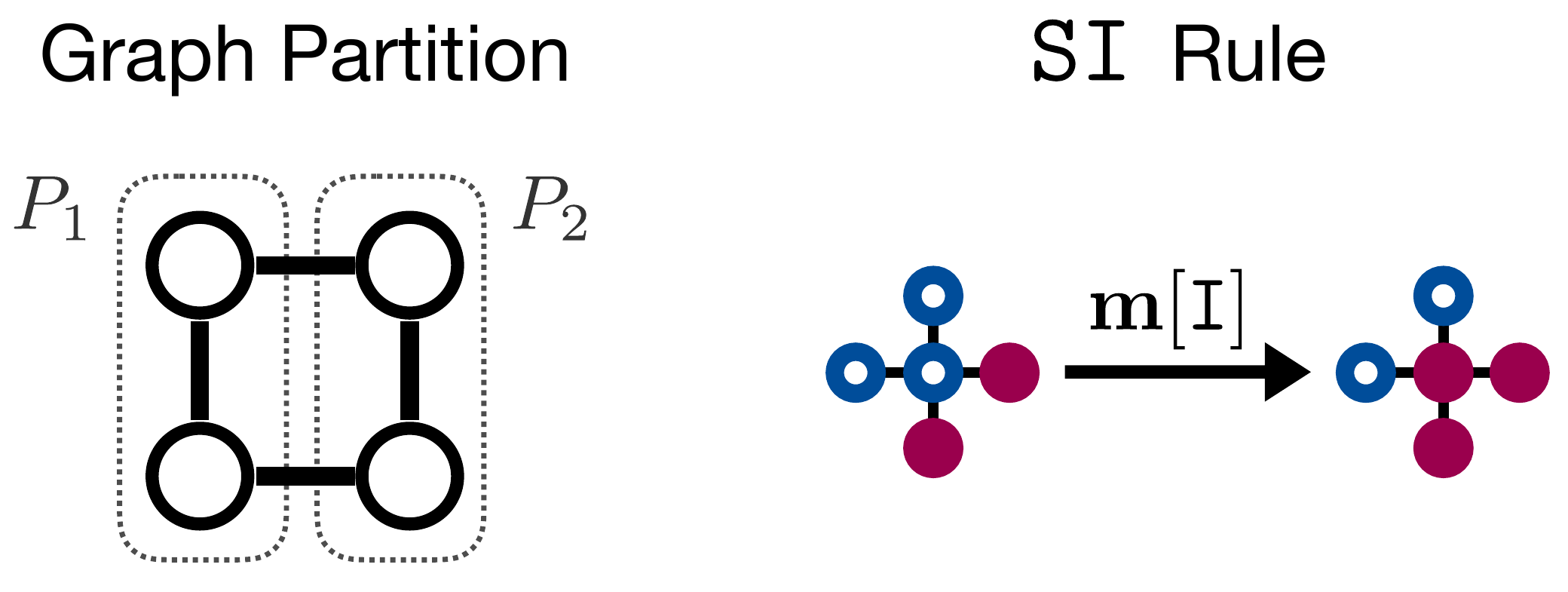}}}%
    \\
    \subfloat[]{{\includegraphics[width=10cm]{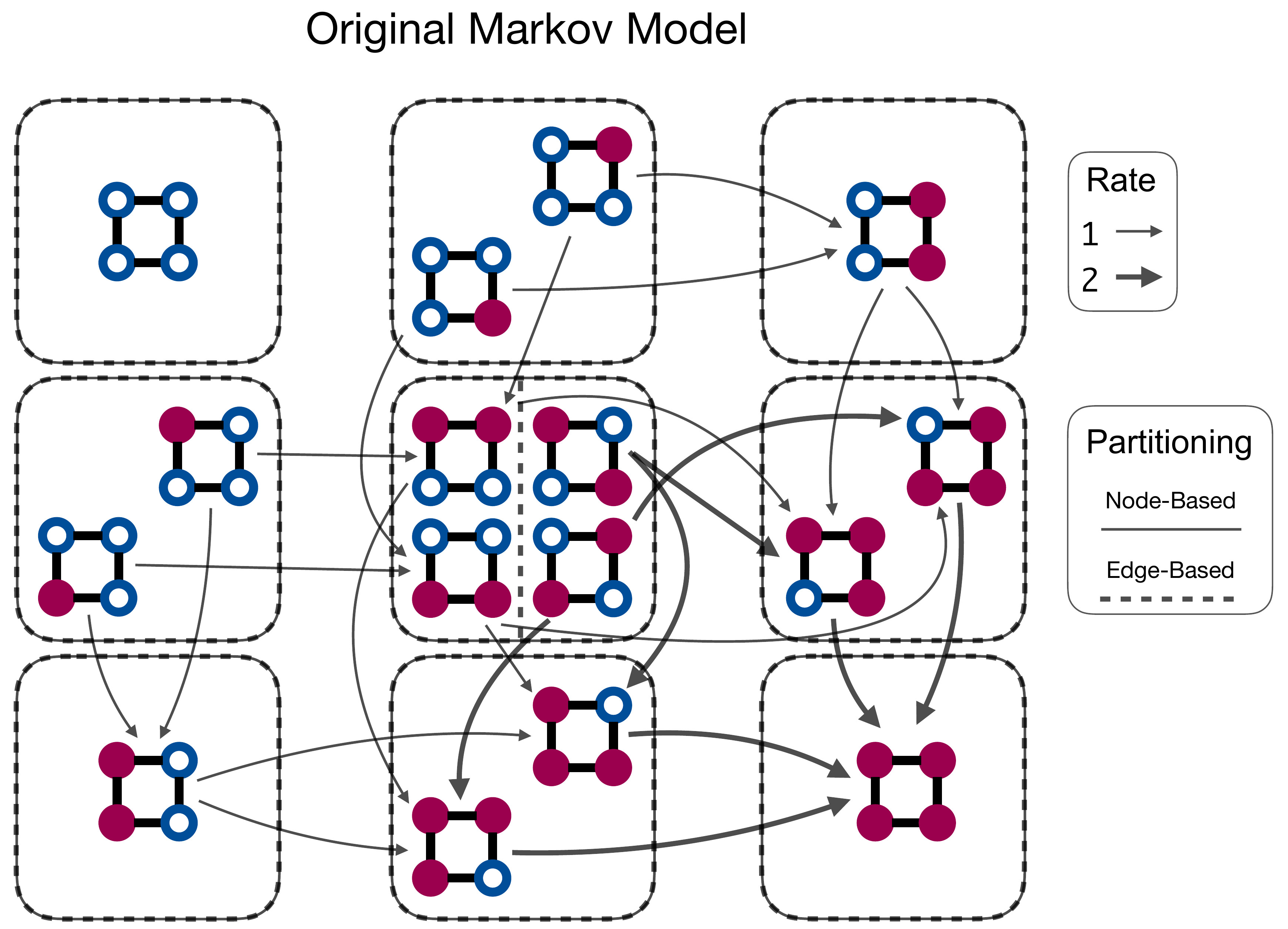} }}%
    \\
    \subfloat[]{{\includegraphics[width=7cm]{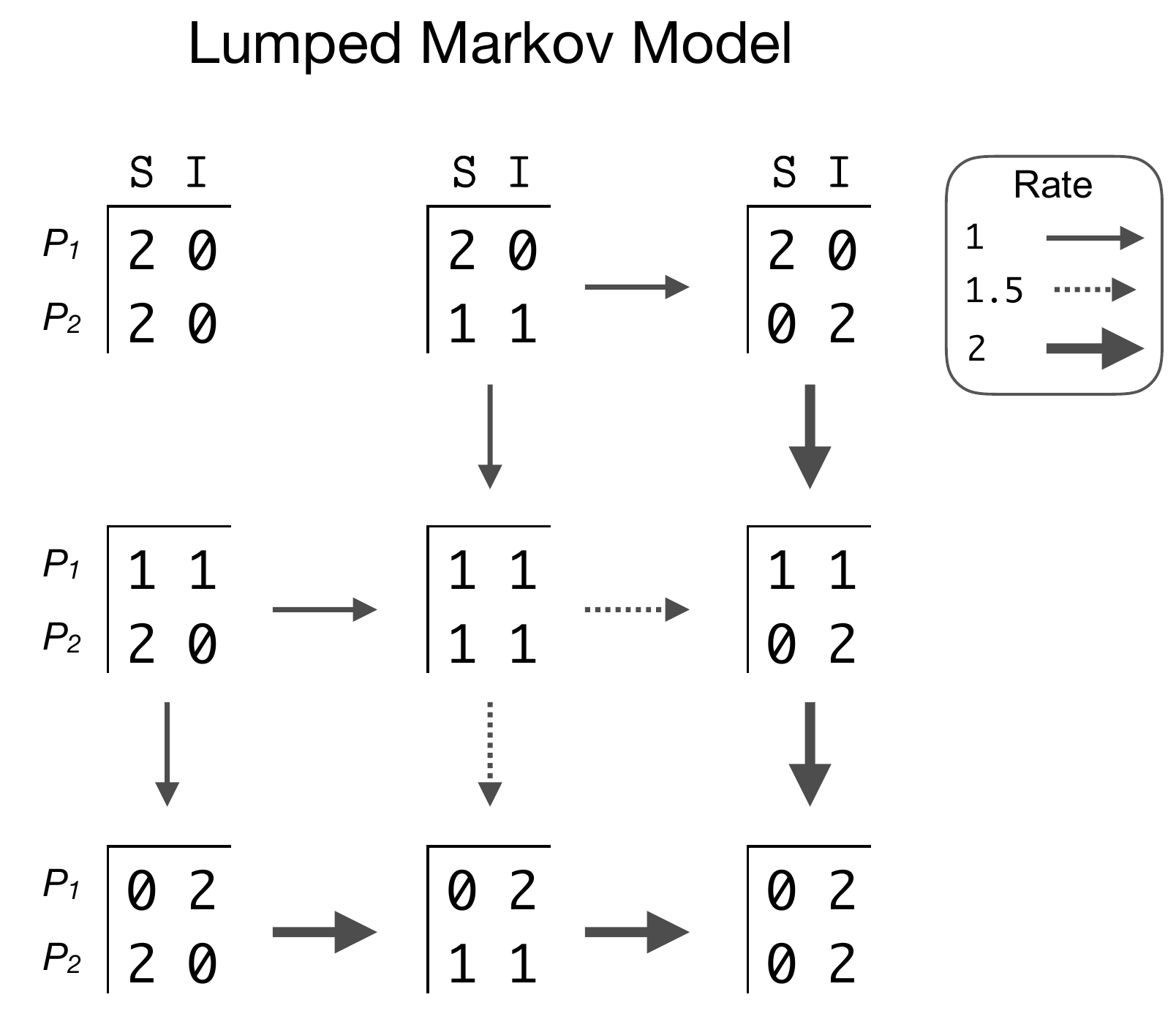} }}%
    \caption{Illustration of the lumping process. 
    (a): Model. A basic $\mathtt{SI}$-Process where infected nodes (magenta, filled) infect susceptible neighbors (blue) with rate infection $\lambda=1$. The contact graph is divided into two partitions. 
    (b): The underlying CTMC with $2^4=16$ states. The graph partition induces the edge-based and node-based lumping. \revision{The edge-based lumping refines the node-based lumping and generates one partition more (vertical line in the central partition).}
    (c): The lumped CTMC using node-based counting abstraction with only 9 states. The rates are the averaged rates from the full CTMC.}%
    \label{fig:example1}%
\end{figure}

\begin{figure}[t]
  \centering
    \subfloat[]{{ \includegraphics[width=0.23\textwidth]{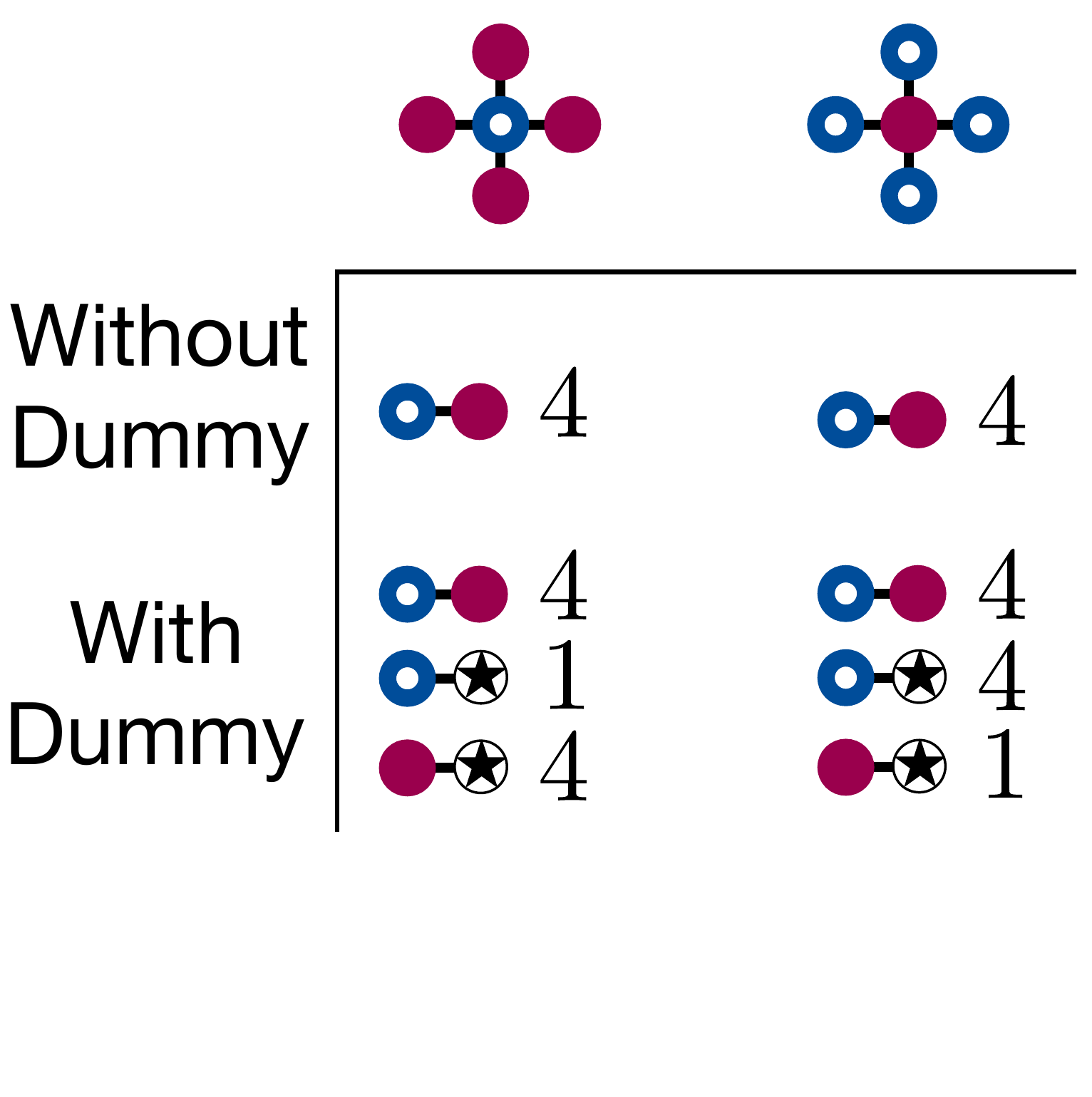}}}%
    \hspace{0.5cm}
    \subfloat[]{{\includegraphics[width=0.3\textwidth]{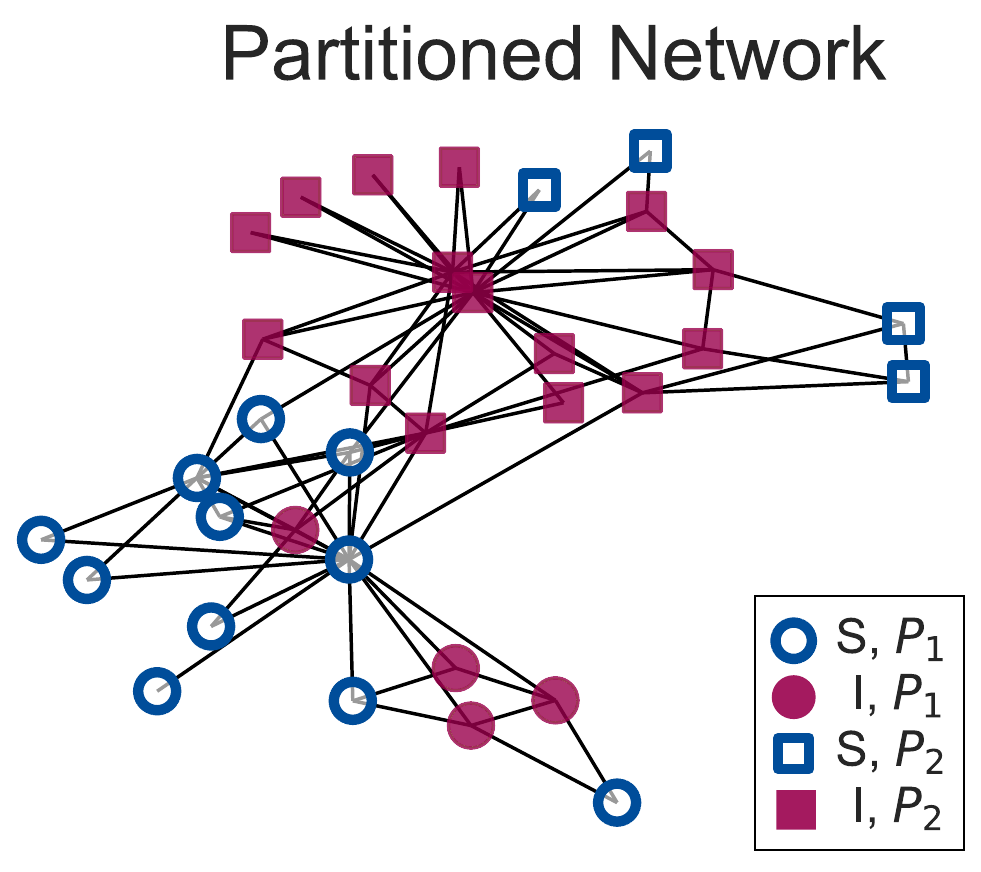}
    \hspace{0.1cm}
     \includegraphics[width=0.34\textwidth]{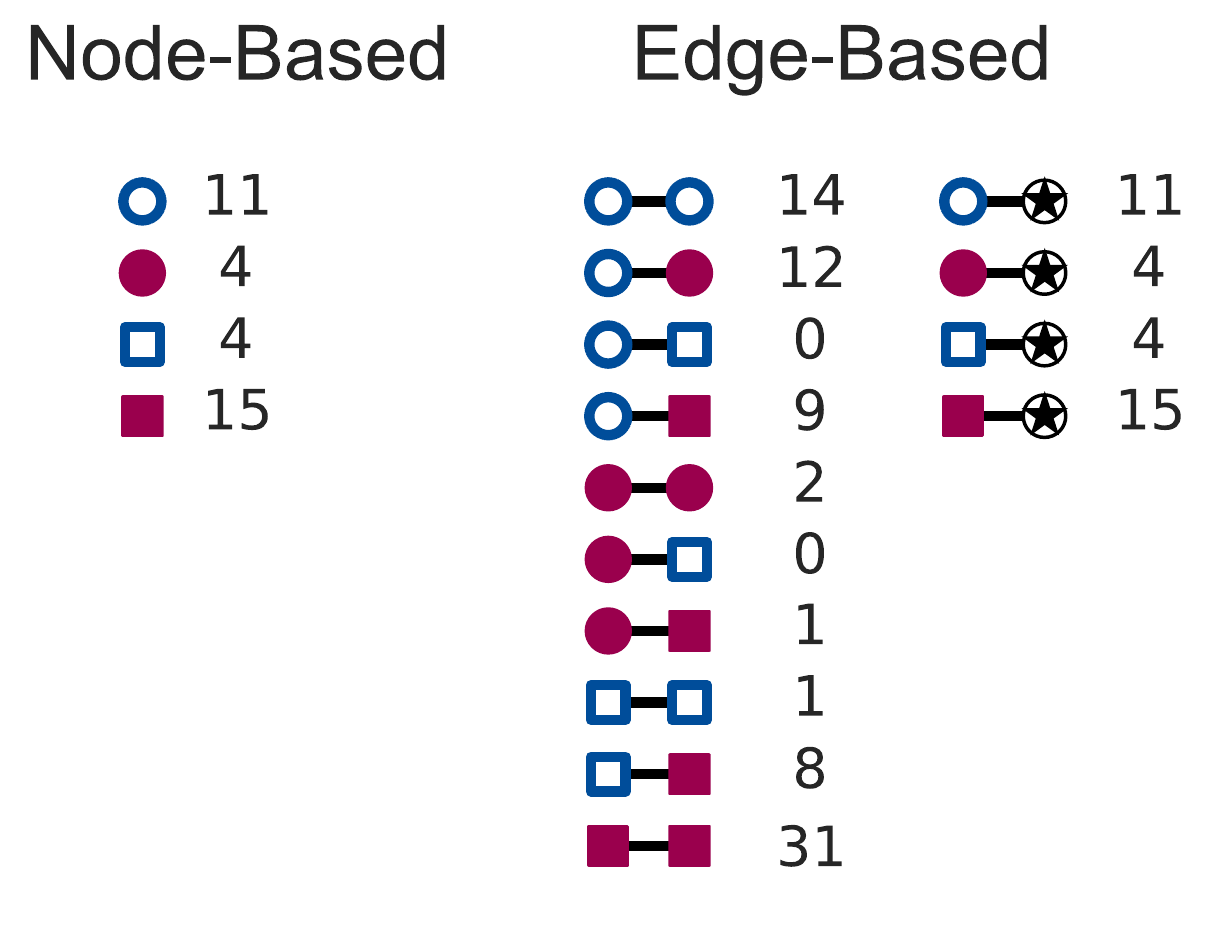}
    }}%
    
  \caption{ \revision{(a) By adding the dummy-node, the edge-based abstraction is able to differentiate the two graphs. 
  Adding the dummy-node ensures that the nodes in each state are counted in the edge-based abstraction. (b) Left: A partitioned network (Zachary's Karate Club graph from \cite{girvan2002community})  ($\texttt{S}$: \emph{blue}, $\texttt{I}$: \emph{magenta, filled}). The network is partitioned into $P_1$ ($\fullmoon$-nodes) and $P_2$ ($\Box$-nodes). Right: The corresponding counting abstractions.}
  }
  \label{fig:dummyNode}
\end{figure}

Next, we construct the lumping function $\mathcal{L}$. Because we want to make our lumping aware of the contact network's topology, we assume a given partitioning $\mathcal{P}$ over the nodes $\mathcal{N}$ of the contact network.
That is,  $\mathcal{P} \subset 2^{\mathcal{N}}$ and $\bigcup_{P \in \mathcal{P}}P = \mathcal{N}$ and all $P \in \mathcal{P}$ are disjoint and non-empty. 
Based on the node partitioning, we can now impose different kinds of counting abstractions on the network state. This work considers two types: counting nodes and counting edges.
The counting abstractions are visualized in Fig.~\ref{fig:dummyNode}. A full example of how a lumped CTMC of an $\mathtt{SI}$ model is constructed using the node-based counting abstraction is given in Fig.~\ref{fig:example1}.

\subsubsection{Node-Based Counting Abstraction}
We count the number of nodes in each state and partition. Thus, for a given network state $x \in \mathcal{X}$, we use $y(s,P)$ to denote the number of nodes in state $s \in \mathcal{S}$ in partition $P\in\mathcal{P}$. The lumping function $\mathcal{L}$ projects $x$ to the corresponding counting abstraction. Formally: 
\begin{align*}
    &\mathcal{Y} = \{y \mid y: \mathcal{S} \times \mathcal{P} \rightarrow \mathbb{Z}_{\geq 0} \} \\ \phantom{\Big(}
    &\mathcal{L}(x) = y   \\
    \text{with:}\hspace{0.5cm}
    &y(s,P) = |\{ n \in \mathcal{N} \mid X(n) = s, n \in P \}| \;.
\end{align*}

\subsubsection{Edge-Based Counting Abstraction}
Again, we assume that a network state $x$ and a node partitioning $\mathcal{P}$ are given.
Now we count the edges, that is for each pair of states $s,s' \in \mathcal{S}$ and each pair of partitions $P,P' \in \mathcal{P}$, we count $y(s,P,s',P')$ which is the number of edges $(n,n') \in \mathcal{E}$
where $x(n)=s$, $n \in P$, $x(n')=s'$, $n' \in P'$. Note that this includes cases where $P=P'$ and $s=s'$. However, only counting the edges does not determine how many nodes there are in each state (see Fig.\;\ref{fig:dummyNode} for an example). 

In order to still have this information encoded in each lumped state, we slightly modify the network structure by adding a new dummy node $n_{\star}$ and connecting each node to it . The dummy node has a dummy state denoted by $\star$ which never changes, and it can be assigned to a new dummy partition $P_{\star}$. \revision{Formally,}
\begin{align*}
&\mathcal{N} := \mathcal{N} \cup \{n_{\star}\} \hspace{0.5cm} 
\mathcal{S} := \mathcal{S}  \cup \{\star\} \hspace{0.5cm}  
L(n_{\star}) = \star \hspace{0.5cm}  
\mathcal{P} := \mathcal{P} \cup \{P_{\star}\}
\\
& \hspace{2.1cm} \mathcal{E} := \mathcal{E} \cup \{(n,n_{\star}) \mid n \in \mathcal{N}, n \neq n_{\star} \} \;.
\end{align*}
Note that the rate function $f$ ignores the dummy node.
The lumped representation is then given as:
\begin{align*}
    &\mathcal{Y} = \{y \mid  y: \mathcal{S} \times \mathcal{P} \times \mathcal{S} \times \mathcal{P} \rightarrow \mathbb{Z}_{\geq 0} \} \\
    &\mathcal{L}(x) = y \phantom{\Big(}  \\
    \text{with:}\hspace{0.5cm}
    &y(s,P,s',P') = |\big\{ (n,n') \in \mathcal{E} \mid x(n) = s, n \in P,  x(n')=s', n' \in P' \big\}| 
\end{align*}

\subsubsection{Example}
Fig.~\ref{fig:example1} illustrates how a given partitioning and the node-based counting approach induces a lumped CTMC. \revision{
The partitions induced by the edge-based counting abstracting are also shown. In this example, the edge-based lumping aggregates only isomorphic network states.
}


\subsection{Graph Partitioning}
Broadly speaking, we have three options to partition the nodes based on local features (e.g., its degree) or global features (e.g.,  communities in the graph) or randomly.    
As a baseline, we use a random node partitioning. Therefore, we fix the number of partitions and randomly assign each node to a partition while enforcing that all partitions have, as far as possible, the same number of elements.

Moreover, we investigate a degree-based partitioning, where we define the distance between to nodes $n,n'$ as their relative degree difference (similar to \cite{kyriakopoulos2018lumping}):
\begin{equation*}
    d_{\text{k}}(n,n') = \frac{| k_n - k_{n'}|}{\text{max}(k_n,k_{n'})} \;.
\end{equation*}
We can then use any reasonable clustering algorithm and build partitions (i.e., clusters) with the distance function. In this work, we focus on bottom-up hierarchical clustering as it provides the most principled way of precisely controlling the number of partitions.
Note that, for the sake of simplicity (in particular, to avoid infinite distances), we only consider contact networks where each node is reachable from every other node. We break ties arbitrarily. 

To get a clustering considering global features we use a spectral embedding of the contract network. Specifically, we use the \url{spectral_layout} function from the \url{NetworkX} Python-package \cite{hagberg2008exploring} with three dimensions and perform hierarchical clustering on the embedding.
In future research, it would be interesting to compute node distances based on more sophisticated graph embedding as the ones proposed in \cite{goyal2018graph}.
Note that in the border cases $|\mathcal{P}|=1$ and $|\mathcal{P}|=|\mathcal{N}|$ all methods yield the same partitioning.

\section{Markov Population Models\label{MPM}}
Markov Population Models (MPMs) are a special form of CTMCs where each CTMC state is a population vector over a set of species. We use $\mathcal{Z}$ to denote the finite set of species (again, with an implicit enumeration) and $\mathbf{y} \in \mathbb{Z}_{\geq 0}^{|\mathcal{Z}|}$ to denote the population vector. Hence, $\mathbf{y}[z]$ identifies the number of entities of species $z$.
The stochastic dynamics of MPMs is typically expressed as a set of reactions $\mathcal{R}$, each reaction, $(\alpha, \mathbf{b}) \in \mathcal{R}$, is comprised of a propensity function $\alpha: \mathbb{Z}_{\geq 0}^{|\mathcal{Z}|}\rightarrow \mathbb{R}_{\geq 0}$ and a change vector $\mathbf{b} \in \mathbb{Z}^{|\mathcal{Z}|}$. When reaction $(\alpha, \mathbf{b})$ is applied, the system moves from state $\mathbf{y}$ to state $\mathbf{y}+\mathbf{b}$. The corresponding rate is given by the propensity function. Therefore, we can rewrite the transition matrix of the CTMC as\footnote{Without loss of generality, we assume that different reactions have different change vectors. If this is not the case, we can merge reactions with the same update by summing their corresponding rate functions.}:
\begin{align*}
q(\mathbf{y},\mathbf{y}') = 
\begin{cases*}
\alpha(\mathbf{y}) & if $\exists (\alpha,\mathbf{b}) \in \mathcal{R}, \mathbf{y}' =  \mathbf{y}+\mathbf{b}$ \\
0       & otherwise
\end{cases*} \;.
\end{align*}

Next, we show that our counting abstractions have a natural interpretation as MPMs.

\subsection{Node-Based Abstraction}
First, we define the set of species $\mathcal{Z}$. Conceptually, species are node states which are aware of their partition:
\begin{equation*}
    \mathcal{Z} = \{ (s,P) \mid s \in \mathcal{S}, P \in \mathcal{P} \} \;.
\end{equation*}
Again, we assume an implicit enumeration of $\mathcal{Z}$. We use $z.s$ and $z.P$ to denote the components of a give species $z$. 

We can now represent the lumped CTMC state as a single population vector $\mathbf{y} \in \mathbb{Z}_{\geq 0}^{|\mathcal{Z}|}$, where $\mathbf{y}[z]$ the number of nodes belonging to species $z$ (i.e., which are in state $z.s$ and partition $z.P$). 
The image of the lumping function $\mathcal{L}$, i.e. the lumped state space $\mathcal{Y}$, is now a subset of
non-negative integer vectors: $\mathcal{Y} \subset \mathbb{Z}_{\geq 0}^{|\mathcal{Z}|}$.

Next, we express the dynamics by a set of reactions.
For each rule $r = s_1 \xrightarrow{f} s_2 $ and each partition $P \in \mathcal{P}$, we define a reaction $(\alpha_{r,P},\mathbf{b}_{r,P})$ with propensity function as:
\begin{align*}
    \alpha_{r,P}:& \mathcal{Y} \rightarrow \mathbb{R}_{\geq 0}  \\
     \alpha_{r,P}(\mathbf{y}) =& \frac{1}{\mathcal{L}^{-1}(\mathbf{y})} \suml{x \in \mathcal{L}^{-1}(\mathbf{y})} \suml{n \in P} 
 f(\vecm_{x,n}) \mathbb{1}_{x(n)=s_1}  \;,
\end{align*}
where $\vecm_{x,n}$ denotes the neighborhood vector of $n$ in network state $x$. Note that this is just the instantiation of Equation \ref{transition} to the MPM framework.

The change vector $\mathbf{b}_{r,P} \in \mathbb{Z}^{|\mathcal{Z}|}$ 
is defined element-wise as:
\begin{align*}
\mathbf{b}_{r,P}[z] = 
\begin{cases*}
1 & if $z.s=s_2,P = z.P$ \\
-1        & if $z.s=s_1,P = z.P$ \\
0       & otherwise
\end{cases*} \;.
\end{align*}
Note that $s_1, s_2$ refer to the current rule and $z.s$ to the entry of $\mathbf{b}_{r,P}$.


\subsection{Edge-Based Counting Abstraction}
We start by defining a \emph{species neighborhood}.
The species neighborhood of a node $n$ is a vector $\mathbf{v} \in \mathbb{Z}_{\geq 0}^{|\mathcal{Z}|}$, where $\mathbf{v}[z]$ denotes the number of neighbors of species $z$. We define $\mathcal{V}_n$ to be the set of possible species neighborhoods for a node $n$, given a fixed contact network and partitioning. 
Note that we still assume that a dummy node is used to encode the number of states in each partition.

Assuming an arbitrary ordering of pairs of states and partitions, we define
\begin{align*}
\mathcal{Z} = \big\{ (s_{source},P_{source},s_{target},P_{target}) \mid & s_{source},s_{target} \in \mathcal{S}, P_{source},P_{target} \in \mathcal{P}, \\ & 
(s_{source},P_{source}) \leq (s_{target},P_{target})  \big\} \;.
\end{align*}

Let us define $\mathcal{V_P}$ to be the set of partition neighborhoods all nodes in $P$ can have:
\begin{equation*}
    \mathcal{V}_P = \bigcup_{n \in P}\mathcal{V}_n \;.
\end{equation*}
For each rule $r = s_1 \xrightarrow{f} s_2 $, and each partition $P \in \mathcal{P}$, and each $\mathbf{v} \in \mathcal{V}_P$, we define a propensity function $\alpha_{r,P,\mathbf{v}}$ with:

\begin{align*}
    \alpha_{r,P,\mathbf{v}}:& \mathcal{Y} \rightarrow \mathbb{R}_{\geq 0}  \\
     \alpha_{r,P,\mathbf{v}}(\mathbf{y}) =& 
     \frac{1}{\mathcal{L}^{-1}(\mathbf{y})}
     \suml{x \in \mathcal{L}^{-1}(\mathbf{y})  } \suml{n \in P} 
 f(\vecm_{x,n}) \mathbb{1}_{x(n)=s_1, V(n)=\mathbf{v}}  \;.
\end{align*}
Note that the propensity does not actually depend on $\mathbf{v}$, it is simply individually defined for each $\mathbf{v}$. The reason for this is that the change vector depends on the a node's species neighborhood. To see this, consider a species $z = (s_{source},P_{source},s_{target},P_{target})$, corresponding to edges connecting a node in state $s_{source}$ and partition $P_{source}$ to a node in state $s_{target}$ and partition $P_{target}$. There are two scenarios in which the corresponding counting variable has to change: (a) when the node changing state due to an application of rule $r$ is the source node, and (b) when it is the target node. Consider case (a); we need to know how many edges are connecting the updated node (which was in state $s_1$ and partition $P$) to a node in state  $s_{target}$ and partition $P_{target}$. This information is stored in the vector  $\mathbf{v}$, specifically in position $\vecv[s_{target},P_{target}]$. The case in which the updated node is the target one is treated symmetrically. This gives rise to the following definition: 
\begin{align*}
\mathbf{b}_{r,P,\mathbf{v}}[z] = 
\begin{cases*}
\vecv[z.s_{target},z.P_{target}]        & if $s_2 = z.s_{source}, P=z.P_{source}$ \\
-\vecv[z.s_{target},z.P_{target}]        & if $s_1 = z.s_{source}, P=z.P_{source}$ \\
\vecv[z.s_{source},z.P_{source}]        & if $s_2 = z.s_{target}, P=z.P_{target}$ \\
-\vecv[z.s_{source},z.P_{source}]        & if $s_1 = z.s_{target}, P=z.P_{target}$ \\
0       & otherwise
\end{cases*} \;.
\end{align*}

The first two lines of the definition handle cases in which the node changing state is the source node, while the following two lines deal with the case in which the node changing state appears as target.

\begin{figure}[t]
  \centering
    \includegraphics[width=0.65\textwidth]{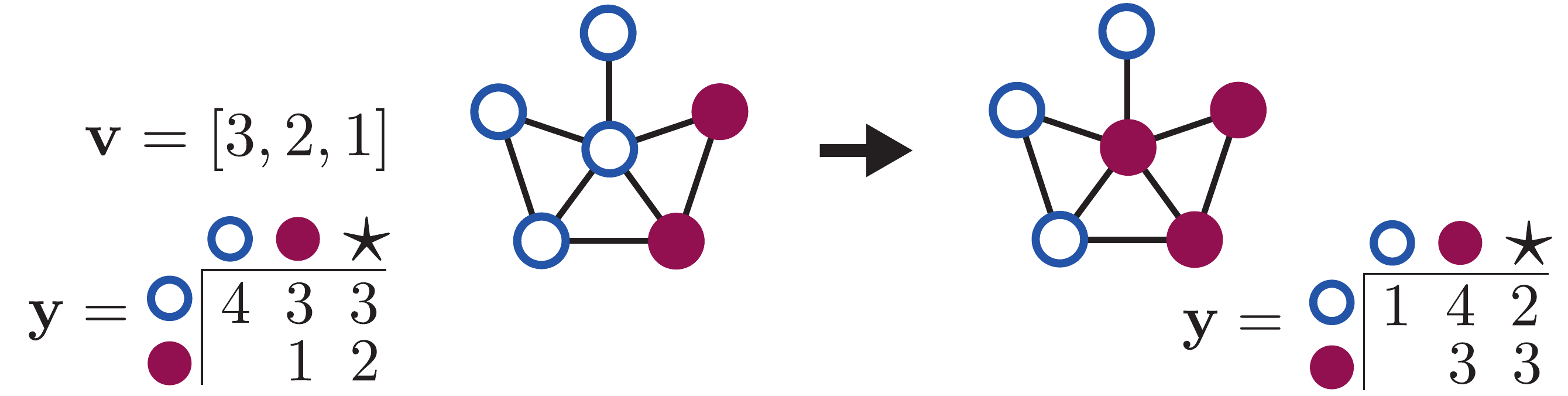}
  \caption{Example of how the neighborhood $\vecv$ influences the update in the edge-based counting abstraction on an example graph. Here, all nodes belong to the same partition (thus, nodes states and species are conceptually the same) and the node states are ordered $[\texttt{S},\texttt{I},\star]$. The population vector $\mathbf{y}$ is given in matrix form for the ease of presentation. 
  }
  \label{edgechange}
\end{figure}

Fig.~\ref{edgechange} illustrates how a lumped network state is influenced by the application of an infection rule.

\subsection{Direct Construction of the MPM}
Approximating the solution of an $\mathtt{SIS}$-type process on a contact network by lumping the CTMC first, already reduces the computational costs by many orders of magnitude. However, this scheme is still only applicable when it is possible to construct the full CTMC in the first place. Recall that the number of network states is exponential in the number of nodes of the contact network, that is,
$|\mathcal{X}| = |\mathcal{S}|^{|\mathcal{N}|} $.

However, in recent years, substantial effort was dedicated to the analysis of very small networks \cite{ward2018general,holme2015shadows,lopez2016stochastic,moslonka2009disease,pautasso2010number}. One reason is that when the size of a network increases, the (macro-scale) dynamics becomes more deterministic because stochastic effects tend to cancel out. For small contact networks, however, methods which capture the full stochastic dynamics of the system, and not only the mean behavior, are of particular importance.

A substantial advantage of the reduction to MPM is the possibility of constructing the lumped CTMC without building the full CTMC first. In particular, this can be done exactly for the node counting abstraction. On the other hand, for the edge counting we need to introduce an extra approximation in the definition of the rate function, roughly speaking introducing an approximate probability distribution over neighboring vectors, as knowing how many nodes have a specific neighboring vector requires us full knowledge of the original CTMC. 
We present full details of such direct construction in Appendix \ref{directconstruction}.

\subsection{Complexity of the MPM}
\begin{thisnote}

The size of the lumped MPM is critical for our method, as it determines which solution techniques are computationally tractable and provides guidelines on how many partitions to choose. There are two notions of size to consider:
(a) the number of population variables and (b) the number of states of the underlying CTMC. While the latter governs the applicability of numerical solutions for CTMCs, the former controls the complexity of a large number of approximate techniques for MPMs, like mean field or moment closure. 

\paragraph{Node-based abstraction.}
In this abstraction, the population vector is of length $|\mathcal{S}|\cdot |\mathcal{P}|$, i.e. there is a variable for each node state and each partition. 

Note that the sum of the population variables for each partition $P$ is $|P|$, the number of nodes in the partition. This allows us to count easily the number of states of the CTMC of the population model: for each partition, we need to subdivide $|P|$ different nodes into $|\mathcal{S}|$ different classes, which can be done in  $\binom{|P|+|\mathcal{S}|-1}{|\mathcal{S}|-1}$ways, giving a number of CTMC states exponential in the number $|\mathcal{S}|$ of node states and $|\mathcal{P}|$ of partitions, but polynomial in the number of nodes:
\begin{equation*}
|\mathcal{Y}| = {\scaleobj{1.2}{\prod_{P \in \mathcal{P}}}} \binom{|P|+|\mathcal{S}|-1}{|\mathcal{S}|-1}\;.
\end{equation*}

\paragraph{Edge-based abstraction.}
The number of population variables, in this case, is one for each edge connecting two different partitions, plus those counting the number of nodes in each partition and each node state, due to the presence of the dummy state. In total, we have $ \frac{q(q-1)}{2} + q$ population variables, with $q=|\mathcal{S}| \cdot |\mathcal{P}|$.


In order to count the number of states of the CTMC in this abstraction, we start by observing that  the sum of all variables for a given pair of partitions $P', P''$ is  the number of edges  connecting such partitions in the graph. 
We use $\epsilon(P',P'')$ to denote the number of edges between $P', P''$ (resp.\;the number of edges inside $P'$ if $P'=P''$). Thus, 

\begin{equation*}
|\mathcal{Y}| \leq {\scaleobj{1.1}{\prod_{\substack{P',P'' \in \mathcal{P}^2 \\ P' \leq P''}}}} \binom{\epsilon(P',P'')+\mathcal{S}^2 -1}{\mathcal{S}^2-1}
\cdot 
{\scaleobj{1.1}{\prod_{P \in \mathcal{P}}}} \binom{|P|+|\oStates|-1}{|\oStates|-1}
\;.
\end{equation*}
This is an over-approximation, because not all combinations are consistent with the graph topology. For example, a high number of infected nodes in a partition might not be consistent with a small number of $\texttt{I}-\texttt{I}$-edges inside the partition. Note that also this upper bound is exponential in  $|\oStates|$ and $|\mathcal{P}|$ but still polynomial in the number of nodes $N$, differently from the original network model, whose state space is exponential in $N$.

The exponential dependency on the  number of species (i.e., dimensions of the population vector) makes the explicit construction of the lumped state space viable only for very small networks with a small number of node states. However, this is typically the case for spreading models like  $\texttt{SIS}$ or $\texttt{SIR}$. Yet, also the number of partitions has to be kept small, particularly in realistic models. We expect that the partitioning is especially useful for networks showing a small number of large-scale homogeneous structures, as happens in  many real-world networks \cite{girvan2002community}.



An alternative strategy for analysis is to derive mean-field \cite{bortolussi2013continuous} or moment closure equations \cite{schnoerr2016} for  MPMs, which can be done without explicitly constructing the lumped (and the original) state space. These are sets of ordinary differential equation (ODE) describing the evolution of (moments of) the population variables.  We refer the reader to \cite{devriendt2017unified} for a similar approach regarding the node-based abstraction. 



\section{Numerical Results\label{casestudies}}


\begin{figure}[t]
    \centering
    \includegraphics[width=0.325\linewidth]{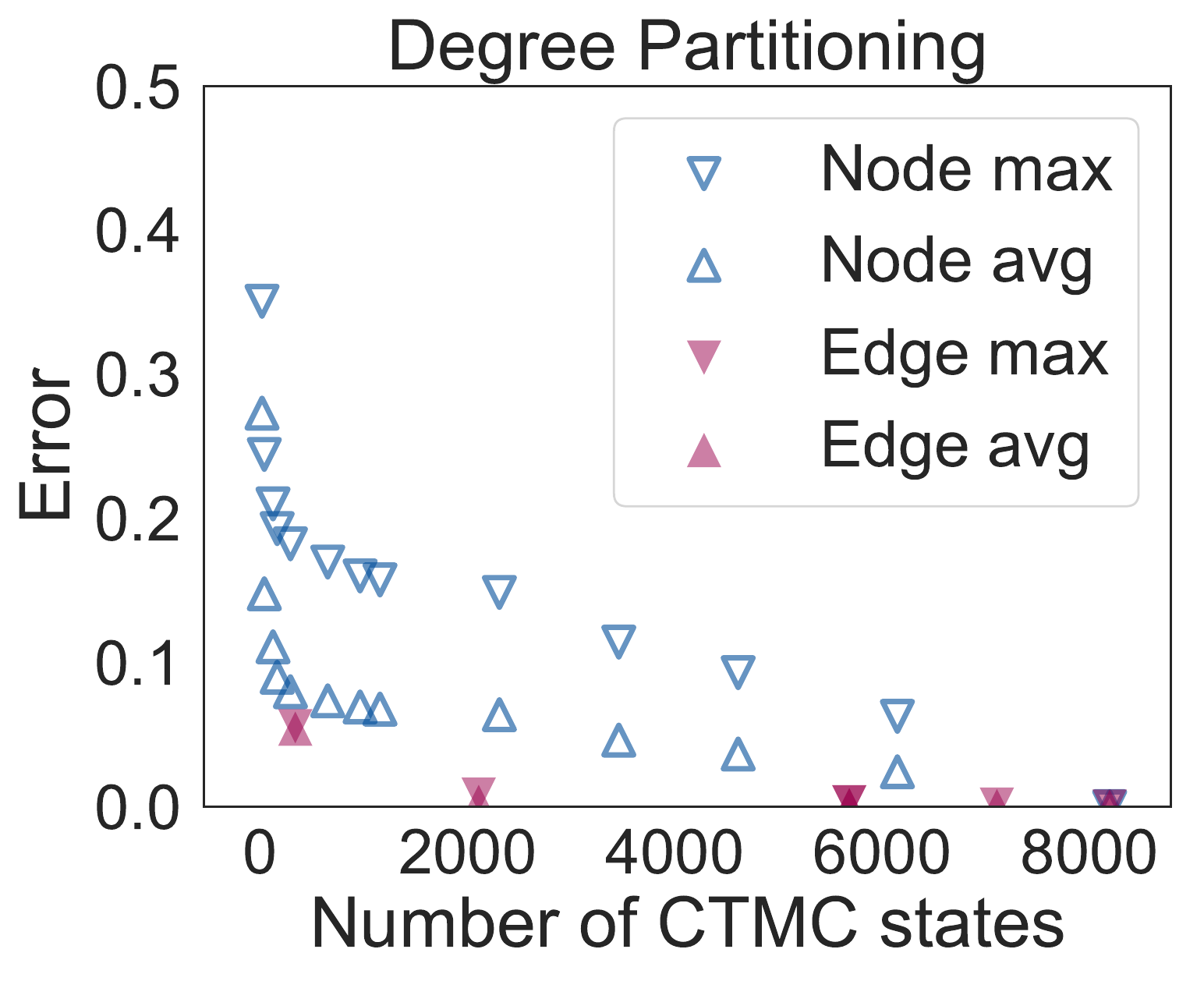}
    \includegraphics[width=0.325\linewidth]{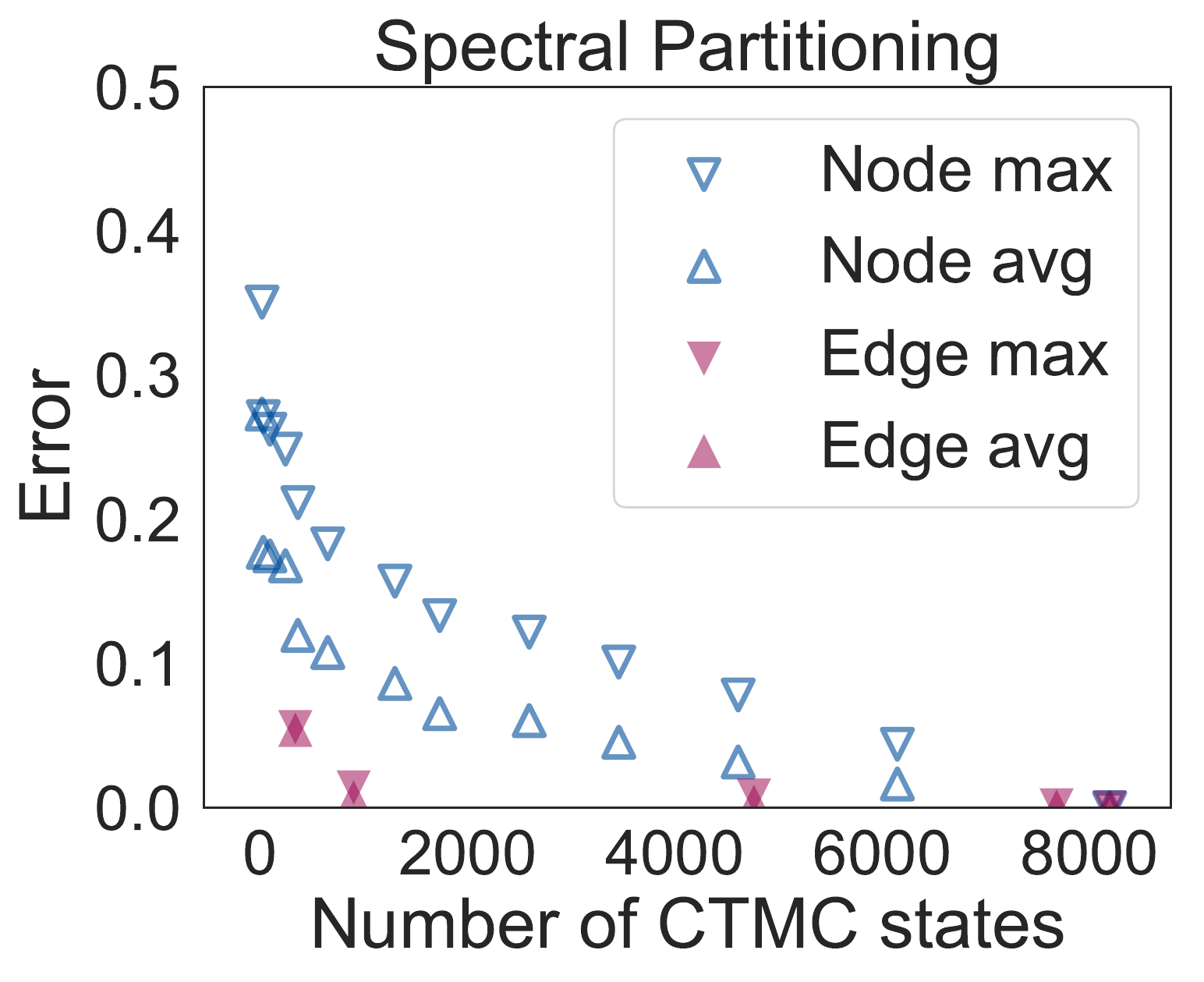}
    \includegraphics[width=0.325\linewidth]{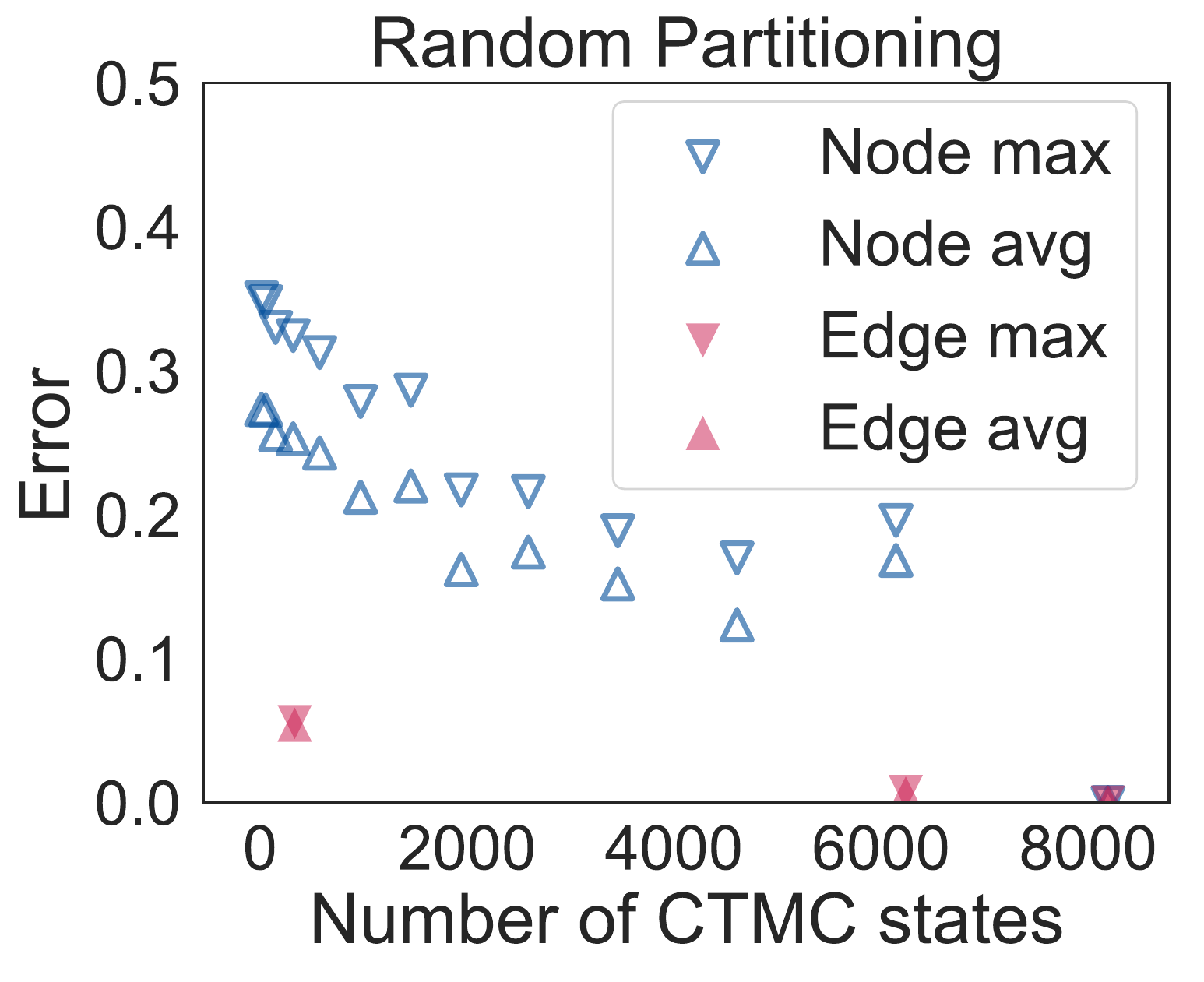}
    \caption{Trade of between accuracy and state space size for the node-based (blue) and edge-based (magenta, filled) counting abstraction. Results are shown for node partitions based on the degree (l.), spectral embedding (c.), and random partitioning (r.). The accuracy is measured as the mean ($\bigtriangleup$) and maximal ($\bigtriangledown$) difference between the original and lumped solution over all timepoints.} 
    \label{results}
\end{figure}

In this section, we compare the numerical solution of the original model---referred to as baseline model---with different lumped MPMs.
The goal of this comparison is to provide evidence supporting the claim that the lumping preserves the dynamics of the original system, with an accuracy increasing with the resolution of the MPM.  
We will perform the comparison by solving numerically the ground and the lumped system, thus comparing the the probability of each state in each point in time.
In practical applications of our method, exact transient or steady state solutions may not be feasible, but in this case we can still rely to approximation methods for MPM \cite{bortolussi2013continuous,schnoerr2016}. Determining which of those techniques performs best in this context is a direction of future exploration. 


A limit of the comparison based on numerical solution of the CTMC is that the state space of the original model has  $|\mathcal{S}|^{|\mathcal{N}|}$ states,  which limits the size of the contact network strongly\footnote{Code is available at \url{github.com/gerritgr/Reducing-Spreading-Processes}}.

Let $P(X(t)=x)$ denote the probability that the baseline CTMC occupies network state $x\in \mathcal{X}$ at time $t\geq 0$. 
Furthermore, let $P(Y(t)=y)$ for $t\geq 0$ and $y \in \mathcal{Y}$ denote the same probability for a lumped MPM (corresponding to a specific partitioning and counting abstraction). To measure their difference, we first approximate the probability distribution of the original model using the lumped solution, invoking the lumping assumption which states that all network states which are lumped together have the same probability mass. We use $P_L$ to denote the \emph{lifted} probability distribution over the original state space given a lumped solution. Formally,
\begin{equation*}
    P_L \big(Y(t)=x\big)=\frac{P\big(Y(t)=y\big)}{|\mathcal{L}^{-1}(y)|} 
    \text{\hspace{0.5cm}} 
    \text{where $y$ is s.t.\;$L(x)=y$.}
\end{equation*}

We measure the difference between the baseline and a lumped solution at a specific time point by summing up the difference in probability mass of each state, then take the maximum error in time:
\begin{equation*}
 d(P,P_L) = \max_t  \sum_{x \in \mathcal{X}} \Big|P_L \big(Y(t)=x)-P(X(t)=x \big) \Big| \;.
\end{equation*}

\end{thisnote}

In our experiments, we used a small toy network with 13 nodes and 2 states ($2^{13} = 8192$ network states). 
We generated a synthetic contact network following the Erdős–Rényi graph model with a connection probability of $0.5$. We use a $\texttt{SIS}$ model with an infection rate of $\lambda=1.0$ and a recovery rate of $\mu=1.3$. Initially, we assign an equal amount of probability mass to all network states.

Fig.~\ref{results} shows the relationship between the error of the lumped MPM, the type of counting abstraction and the method used for node partitioning. We also report the mean difference together with the maximal difference over time.



From our results, we conclude that the edge-based counting abstraction yields a significantly better trade-off between state space size and accuracy. However, it generates larger MPM models than the node-based abstraction when adding a new partition.
We also find that spectral and degree-based partitioning yield similar results for the same number of CTMC states and that random partitioning performed noticeably worse, for both edge-based and node-based counting abstractions. 


\section{Conclusions and Future Work\label{conclusion}}
This work developed first steps in a unification of the analysis of stochastic spreading processes on networks and Markov population models. 
Since the so obtained MPM can become very large in terms of species, it is important to be able to control the trade-off between state space size and accuracy.

However, there are still many open research problems ahead. 
Most evidently, it remains to be determined which of the many techniques developed for the analysis of MPMs (e.g. linear noise, moment closure) work best on our proposed epidemic-type MPMs and how they scale with increasing size of the contact network. 
We expect also that these reduction methods can provide a  good starting point for deriving advanced mean-field equations, similar to ones in \cite{devriendt2017unified}.
Moreover, literature is very rich in proposed moment-closure-based approximation techniques for MPMs, which can now be utilized \cite{soltani2015conditional,grima2012study}.
We also plan to investigate the relationship between lumped mean-field equations \cite{grossmann2018lumping,kyriakopoulos2018lumping} and coarse-grained counting abstractions further. 

Future work can additionally explore counting abstraction of different types, for instance, a neighborhood-based abstraction like the one proposed by James P.~Gleeson in \cite{gleeson2011high,gleeson2013binary}.

Finally, we expect that there are many more possibilities of partitioning the contact network that remain to be investigated and which might have a significant impact on the final accuracy of the abstraction. 

\subsubsection{Acknowledgements}
This research has been partially funded by the German Research Council (DFG)
as part of the  Collaborative Research Center
\say{Methods and Tools for Understanding and Controlling Privacy}.
We thank Verena Wolf for helpful discussions and provision of expertise. 

\newpage
\bibliographystyle{abbrv}
\bibliography{mylib2.bib}


\newpage
\section*{Appendix}
\addcontentsline{toc}{section}{Appendix}

\section{Direct Construction of MPMs\label{directconstruction}}
Here, we prosper a way of directly deriving the lumped MPMs from the contact network without building the original CTMC first. We start with the node-based counting abstraction.

\subsection{Node-Based Abstraction with General Rate Functions}
Our general strategy is to iterate over the nodes in the contact network and to compute the mean rate attributed to that node over all $x\in\mathcal{X}$. Therefore, we consider the possible states of each node together with all possible species neighborhoods. The probability of a node $n$ being in state $s$ and having species neighborhood $\mathbf{v}$ is denoted as $\Pr \big (X(n)=s,V(n)=\vecv \big)$.

For a specific rule $r=s_1 \xrightarrow{f} s_2$ and partition $P$, we can then describe $\alpha_{r,P}$ as:
\begin{align*}
\alpha_{r,P}(\mathbf{y}) = 
\suml{n \in P} \phantom{.} \suml{\vecv \in \Vecv_n}  f(\vecm_{\vecv}) \Pr \big (X(n)=s_1,V(n)=\vecv \big) \;,
\end{align*}
where $\vecm_{\vecv}$ is the neighborhood vector induced by $\vecv$, which we receive by grouping all partitions together. 
Note that it is not computationally necessary to actually iterate over all nodes in the partition. Instead we can group all nodes with the same partition neighborhood together, that is, all nodes $n',n''\in P$ with $V_{n'} = V_{n''}$ as the probability only depends on $\vecv$.

Computing the probability is the interesting part, we start by establishing that 
\begin{align*}
   & \Pr \Big (X(n)=s_1,V(n)=\vecv \Big) \\
   = & \Pr \Big (X(n)=s_1 \Big) \cdot \Pr \Big (V(n)=\vecv \mid X(n)=s_1 \Big)  \;.
\end{align*}
The first term in the product can be described by simply dividing the number of $s_1$-nodes in $P$ with the total number of nodes in $P$.
\begin{align*}
   \Pr \Big (X(n)=s_1 \Big) = \frac{\mathbf{y}[s_1, P]}{|P|} \hspace{0.5cm} \text{where: }n \in P \;.
\end{align*}
The latter probability can be computed for each partition independently. This is because we know the number of nodes in each state in each partition. 
We also know that in partition $P$ the current node $n$ is already in state $s_1$, which we have to take into account. First, we define $\mathbf{y}_P \in \mathbb{Z}_{\geq 0}^{|\mathcal{S}|}$ to the the projection from $y$ to $P$. Thus, each entry is defined by:
\begin{align*}
 \mathbf{y}_P[s] = \mathbf{y}[s,P] \;.
\end{align*}
Likewise, we define $\mathbf{v}_P \in \mathbb{Z}_{\geq 0}^{|\mathcal{S}|}$, such that 
$\mathbf{v}_P[s] = \mathbf{v}[s,P]$. We also define $V(n)_P \in \mathbb{Z}_{\geq 0}^{|\mathcal{S}|}$ to be the number of neighbors of node $n$ in partition $P$ for each state.
Finally, we define $\mathbf{y}_P^{s_1^-}$ to be the same vector as $\mathbf{y}_P$ except that the entry corresponding to state $s_1$ is subtracted by one (and truncated at zero). 
We can now rewrite the probability as:
\begin{align*}
&  \Pr \Big (V(n)=\vecv \mid X(n)=s_1 \Big) \\
  =&  \prod_{P' \in \mathcal{P}} \Pr \Big(V(n)_{P'}=\mathbf{v}_{P'}\bigm| X(n)=s_1 \Big)  \\
   =&  \Pr \Big(V(n)_{P}=\mathbf{v}_{P} \bigm| X(n)=s_1 \Big)
   \prod_{P' \in \mathcal{P} \setminus \{P \}} \Pr \Big(V(n)_{P'}=\mathbf{v}_{P'} \bigm| X(n)=s_1 \Big)  \\
 =& p_h \Big(\mathbf{v}_{P} ; \mathbf{y}_P^{s_1^-} \Big)  
 \cdot 
 \prod_{P' \in \mathcal{P} \setminus \{P \}} p_h \Big(\mathbf{v}_{P'} ; \mathbf{y}_{P'} \Big) \;.
\end{align*}

We use $p_h(\mathbf{k} ; \mathbf{K})$ to denote the probability mass function of the the multivariate hypergeometric distribution, where $\mathbf{k}, \mathbf{K}$ denote to vectors over non-negative integers of the same length. That is, if $\mathbf{K}$ denotes the number of nodes in each state in a partition (resp., the number of marbles in an urn with different colors), then, $p_h(\mathbf{k} ; \mathbf{K})$ denotes the probability of drawing exactly $\mathbf{k}[s]$ nodes (resp. marbles) of each state (resp. color).

\subsection{Reaction Networks and Linear Models}
A special case of MPMs are biochemical reaction networks, where the species represent different types of molecules. The change vectors and corresponding propensity functions can elegantly be expressed as monomolecular ($\texttt{A} \rightarrow \texttt{B}$) and bimolecular ($\texttt{A}+\texttt{B} \rightarrow \texttt{C}+\texttt{D}$) reaction rules ($\texttt{A},\texttt{B},\texttt{C},\texttt{D} \in \mathcal{Z}$). 

\subsubsection{Reduction to Biochemical Reaction Networks}
Most classical models in computational epidemiology are solely comprised of \emph{node-based} rules (like the curing rule) and \emph{edge-based} rules (like the infecting propagation rule). We call these \emph{linear models}.
Node-based rules, also referred to as \emph{spontaneous} or \emph{independent} rules, have a constant rate function, i.e., $f(\vecm)=\mu$. Edge-based rules, also referred to as \emph{contact} rules, are linear in exactly one dimension, i.e., they have the form $f(\vecm)=\lambda \vecm[s]$. 

Linear models are special because not the whole neighborhood is important for the rate of a rule but only the expected number of neighbors in a certain state. This makes the rule very similar to monomolecular and bimolecular reaction rates in MPMs. In fact, we can model the whole dynamics as a set of reaction over the species $\mathcal{Z}$.

Chemical reaction networks are a special case of Markov population models.
In a chemical reaction network the state space is given by population vectors over species and molecular reactions have the form $\texttt{A}\xrightarrow{a}\texttt{C}$ or $ \texttt{A}+\texttt{B}\xrightarrow{b}\texttt{C}+\texttt{D}$, where $\texttt{A,B,C,D}$ denote species and $a,b \in \mathbb{R}_{\geq 0}$ are reaction rate constants.

For each node-based rule $s_1 \xrightarrow{\mu} s_2$, we construct the reactions
\begin{equation*}
    (s_1,P) \xrightarrow{\mu } (s_2,P) \hspace{1cm} \forall P \in \mathcal{P}
\end{equation*}
For each edge-based rule $s_1 \xrightarrow{f} s_1$, $f(\vecm)=\lambda \vecm[s']$, we construct the reactions
\begin{equation*}
    (s_1,P) + (s',P')  \xrightarrow{\lambda w_{P,P'} } (s_2,P) + (s',P') \hspace{1cm} \forall P,P' \in \mathcal{P}
\end{equation*}

where $w_{P,P'}$ denotes the mean number of edges of a random node in $P$ with nodes in $P'$, that is\footnote{Note that, despite the duple notation, we only count edges once}:
\begin{align*}
w_{P,P'} = 
\begin{cases*}
\frac{\epsilon(P,P)}{|P|}  \frac{1}{|P|-1}  & if $P=P'$ \\
 \frac{\epsilon(P,P')}{|P|} \frac{1}{|P'|}      & otherwise
\end{cases*} \;.
\end{align*}
with
\begin{equation}
\epsilon(P,P') = | \{(n_1,n_2) \in \mathcal{E} \mid n_1 \in P, n_2 \in P' \} | \;.
\end{equation}

\subsection{Edge-Based Counting Abstraction}
For each rule $r = s_1 \xrightarrow{f} s_2 $, and each partition $P \in \mathcal{P}$, and each $\mathbf{v} \in \mathcal{V}_P$, we define a propensity function $\alpha_{r,P,\vecv}$ with:
\begin{align*}
\alpha_{r,P,\mathbf{v}}(\mathbf{y}) = 
\suml{n \in P}  f(\vecm_{\vecv}) \Pr \big (X(n)=s_1,V(n)=\vecv \big) \;.
\end{align*}
Again, we use 
\begin{align*}
\Pr \big (X(n)=s_1,V(n)=\vecv \big)
   = \Pr \big (X(n)=s_1 \big) \cdot \Pr \big (V(n)=\vecv \mid X(n)=s_1 \big) 
\end{align*}
to compute this probability, where we can solve $\Pr \big (X(n)=s_1 \big) $ exactly as before. 

Since we have now information about the edges, we can derive the probability of neighborhoods more precisely. In fact, we can directly construct the set of candidate neighbors from $\mathbf{y}$. Therefore, we define a vector $\mathbf{y}_{s,P,P'} \in \mathbb{Z}_{\geq 0}^{|\mathcal{S}|}$, where entry $\mathbf{y}_{s,P,P'}[s']$ specifies the number of neighbors of a random node in state $s$ and partition $P$, which lie in partition $P'$ and occupy state $s'$.
Formally:

\begin{align*}
\mathbf{y}_{s,P,P'}[s'] =
    \begin{cases*}
      \mathbf{y}[s,P,s',P'] & if $(s,P) \leq (s',P')$ \\
     \mathbf{y}[s',P',s,P]        & otherwise
    \end{cases*}
    \;.
\end{align*}


This gives rise to the final approximation of the probability of neighborhood species: 
\begin{align*}
 \Pr \Big (V(n)=\vecv  \bigm|  X(n)=s_1 \Big) \approx \prod_{P' \in \mathcal{P}} p_h \big(\vecv_P ; \mathbf{y}_{s_1,P,P'} \big)  \;\;\; (\text{where } n \in P.)
\end{align*}

\newpage

\newpage

\end{document}